\documentclass{fmj2010}

\def\fermilat{\textit{Fermi}/LAT}

\def\XMM{\textit{XMM-Newton }}
\def\Cha{\textit{Chandra }}
\begin{document}
   \title{Young radio sources: a radio-gamma perspective}

   \author{M. Orienti\inst{1,2}
          \and
          G. Migliori\inst{3}
          \and
          A. Siemiginowska\inst{4}
          \and
          A. Celotti\inst{3}
          }

   \institute{Dipartimento di Astronomia, Universit\'a di Bologna, via
     Ranzani 1, I-40127, Bologna, Italy 
         \and
             Istituto di Radioastronomia - INAF, via Gobetti 101,
             I-40129, Bologna, Italy
          \and
             SISSA/ISAS, via Bonomea 265, I-34136, Trieste, Italy
          \and Harvard-Smithsonian Center for Astrophysics, 60 Garden St, Cambridge, MA 02138, USA
             }

   \abstract{
The evolutionary stage of a powerful radio source originated by an AGN
is related to its linear size. In this context, compact symmetric
objects (CSOs), which are powerful and intrinsically small ($<$ 1 kpc)
radio sources with a convex synchrotron radio spectrum that peaks
around the GHz regime, should represent a young stage in the
individual radio source life. Their radio jets expand within the dense
and inhomogeneous interstellar medium of the host galaxy, which may
influence the source growth. The radio emission is expected to evolve
as a consequence of adiabatic expansion and radiative and inverse
Compton losses. The role played by the different mechanisms in the
radio and gamma regimes is discussed.  
   }

   \maketitle
%

\section{Introduction}

Powerful (L$_{\rm 1.4 GHz}$ $>$ 10$^{25}$ W/Hz) and intrinsically
compact (linear size LS $<$ 1-20 kpc) extragalactic radio sources are
considered to represent an early stage in the individual radio source
evolution. Their main characteristic is the rising synchrotron radio spectrum
which turns over at frequencies from a few hundred MHz up to the GHz
regime. The mechanism responsible for the spectral peak is the
synchrotron self-absorption (SSA; Snellen et al. \cite{snellen00}),
although an additional contribution from free-free absorption (FFA) is
present in the most compact objects (Orienti \& Dallacasa \cite{mo08},
Kameno et al. \cite{kameno00}).\\
When imaged with the high spatial resolution provided by radio
interferometers, these objects resemble a scaled-down version of the
classical edge-brightened FR II radio galaxies (Fanaroff \& Riley
\cite{fr74}). Given their compact and two-sided structures,
Wilkinson et al. (\cite{wilkinson94}) termed these sources
``Compact Symmetric Objects'' (CSO) and suggested a possible
evolutionary connection with the larger radio galaxies. 
Conclusive evidence of the genuine {\it youth} of this class of
objects came from the determination of both kinematic 
(Polatidis \& Conway \cite{pc03}) and radiative (Murgia
\cite{murgia03}, Murgia et al. \cite{murgia99}) ages, which resulted
to be in the range of 10$^{3}$
- 10$^{5}$ years.  \\ 
Several studies of samples of compact radio sources (O'Dea \&
Baum \cite{odea97}) show an anti-correlation between the peak
frequency and the linear size: the higher the peak frequency $\nu_{\rm
p}$, the smaller the
source is. This anti-correlation agrees well 
with an evolutionary scenario where the peak, caused by
SSA, moves to lower frequencies as the source adiabatically
expands. In this context we can draw an evolutionary path linking the
different stages of the source evolution: the smaller sources (LS $<$ 1
- 50 pc), with $\nu_{\rm p}$ above a few GHz and known as high
frequency peakers (HFP), will evolve into the GHz-peaked spectrum (GPS)
sources, with LS $\sim$ 1 kpc and $\nu_{\rm p} \sim$ 1 GHz, which will
become compact steep spectrum (CSS) objects (LS $\sim$ 1 - 20 kpc, and
$\nu \sim$ 100 MHz), i.e. the progenitors of FRIIs.\\ 
Several evolutionary models (i.e Fanti et al. \cite{cf95}, Kaiser \&
Alexander \cite{kaiser97}) have been developed to describe the
various steps of the source growth. Given the compact sizes, these
objects entirely reside within the host galaxy, enshrouded by the
dense and inhomogeneous interstellar medium (ISM), where jet-cloud interaction
may play a role on the source growth (Jeyakumar \cite{jeyakumar09}).\\
In this contribution we describe the role played by the various
mechanisms on the source evolution, such as adiabatic expansion,
energy losses, and their effects on the source spectrum, in particular
in the radio and gamma regimes.\\

Throughout this paper, we assume 
$H_{0} = 71$ km s$^{-1}$ Mpc$^{-1}$, $\Omega_{\rm M} = 0.27$,
$\Omega_{\Lambda} = 0.73$, in a flat Universe. The spectral index is
defined as $S$($\nu$) $\propto \nu^{- \alpha}$. \\ 

\section{Physical properties}

The knowledge of the magnetic field has crucial implication in the
determination of the radio source evolution. The models developed so
far consider that the radio sources are in minimum energy condition,
which corresponds to equipartition between the particle energy and the
magnetic field (Pacholczyk \cite{pacho70}). However, there is no {\it
  a-priori} reason why a radio source must be in equipartition. \\
A direct way to measure the magnetic field is based on observational
parameters. In fact, in the case the spectral peak is produced by SSA,
the magnetic field $H$ can be computed from peak frequency $\nu_{\rm p}$,
peak flux density $S_{\rm p}$, and angular size $\theta$ by the
relationship (Kellermann \& Paulinity-Toth \cite{paulini81}):

\begin{equation}
H \sim f(\alpha) \theta^{4} \nu_{\rm p}^{5} S_{\rm p}^{-2} {\rm
  (1+z)^{-1}}
\label{ssa}
\end{equation}

\noindent where $f(\alpha)$ is a function weakly dependent on the
spectral index
($f_{\alpha}$=8 for $\alpha$=0.5), and $z$ is the redshift.
The region considered in Eq. \ref{ssa} is assumed to be homogeneous
and deviations from this assumption introduce uncertainties. Studies of
a few CSS carried out at low frequencies (i.e. close to their peak
frequency) were subjected to substantial uncertainties on the component
size. The high spatial resolution and the frequency coverage of the
VLBA match well the requirements for a proper study of the most
compact HFP sources, whose spectrum is reasonably fitted by
self-absorbed synchrotron emission from a homogeneous component.
The magnetic field derived for a sample of very young and compact
HFP (Orienti \& Dallacasa \cite{mo08}) has been found 
in good agreement with the
equipartition magnetic field, with values around a few tens of mG.\\
In the presence of such high magnetic fields, the radiative losses are
very severe, making the lifetime of the relativistic electrons
responsible for the radio emission very short, and causing a marked
cut-off in the optically-thin part of the spectrum. In the case the
particle injection stops, the radio spectrum rapidly shifts
towards low frequencies becoming 
undetectable by conventional observations, 
since only low-energy
electrons will be able to survive longer in such high magnetic fields.\\
It should be noted that in a few source components, 
the magnetic field computed directly from
the observational parameters are very different from the equipartition
value. In these cases, the analysis of their radio spectra indicates
that the optically-thick regime is too inverted to be due to SSA
(see Section 3), and an
additional contribution from FFA is needed. This implies that the
magnetic field derived from observational parameters is physically
meaningless.\\

\section{The ambient medium}

The onset of the radio emission is thought to be related to merger or
accretion events which feed the central active galactic nucleus
(AGN). For this reason, the interstellar medium of galaxies hosting a
radio source is rather dense and inhomogeneous. Statistical
studies of the atomic hydrogen in absorption of a sample of young
radio sources (Pihlstr\"om et al. \cite{ylva03}, Gupta et
al. \cite{gupta06}) have shown an anti-correlation between the linear
size and the HI column density ($N_{\rm HI}$): the larger the source, the
smaller the HI column density is. This can be explained assuming
that the neutral hydrogen is settled in a circumnuclear torus/disk:
the HI absorption is detected against the receding jet when our line
of sight passes through the disk/torus along its way toward the radio
emission. This interpretation is based on observations with a poor
spatial resolution. When pc-scale VLBI observations have been
performed, it has been noted that in some compact sources, like
4C\,12.50 (Morganti et al. \cite{morganti04}) the HI absorption is not
produced by an organized circumnuclear structure, but it probably comes from an
unsettled off-nuclear cloud located where the jet bends.\\
Another evidence of an inhomogeneous medium enshrouding the radio
sources arises from the distribution of the ionized medium. VLBI
studies of the compact HFP sources OQ\,208 (Kameno et
al. \cite{kameno00}), J0428+3259, and J1511+0518 
(Orienti \& Dallacasa \cite{mo08})
could locate FFA against one lobe only, indicating an
asymmetric distribution of the ionized gas.
Given their
intrinsically small linear size, young radio sources completely reside
within such a dense and inhomogeneous environment, 
where jet-ISM interaction may
influence the source evolution. Various work on young radio sources
pointed out that in a large number of objects the brightest lobe is
also the closest to the core (Saikia et al. \cite{saikia03}, Orienti
et al. \cite{mo07}). Such asymmetries are better interpreted in terms
of jet-ISM interaction, instead of projection effects. 
Furthermore, the detection of HI in absorption only against 
the brightest (and closest to core) lobe of 3C\,49 and 3C\,268.3 (Labiano et
al. \cite{labiano06}), strongly supports this interpretation. 
During the time
the jet is piercing the cloud, its velocity is considerably slowed
down. The interaction prevents adiabatic expansion and the radio
luminosity is enhanced by strong radiative losses.\\

\section{High energy emission}

Although young radio sources are preferentially studied in the radio
band, the knowledge of their high energy emission is crucial in
providing us information on the central region of the AGN and on 
their total energy distribution, i.e. a key element in understanding
the source fate.\\ 
X-ray detections of GPS/CSSs registered a drastic increase with the
advent of \XMM and \Cha observatories. Observational campaigns on  GPS
and CSS galaxies 
(Guainazzi et al. \cite{Gua06}, Vink et al. \cite{Vin06},
Tengstrand et al. \cite{Ten09}) and quasars (Siemiginowska et al. \cite{Sie08}) 
have been performed for the first time with detection 
fractions nearly to the 100\% on the selected subsamples. 
However, for most of the cases, the extreme compactness of the sources 
combined with the spatial resolution of the X-ray observatories
($\sim1''$ in the best case with $Chandra$) prevents to resolve out 
the X-ray morphology and locate the site of origin of the high-energy emission.
Therefore, the studies rely mainly on the analysis of 
the X-ray spectral properties (presence of an intrinsic absorber, 
evaluation of X-ray intrinsic luminosities), but the identification
of the various spectral features is often controversial. 
As a consequence, the origin of the X-ray and high-energy emission 
remains still a matter of debate.
Thermal high energy/X-ray emission is expected from 
the accretion disk's hot corona or it could arise from interactions 
between the expanding radio source and the interstellar medium 
(Reynolds et al. \cite{RHB01}, Bicknell \& Sutherland \cite{BS06}). 
A significant contribution to the total X-ray flux can be given 
by the extended components, namely jet, hot spots and lobes, 
overpressured and powerful at the initial stages.
In this case, the mechanism at the basis of high energy emission
may be inverse Compton (IC) of either the thermal UV/IR photons from
the accretion disk/circumnuclear torus by the lobes' relativistic
electrons (Stawarz et al. \cite{stawarz08}, Ostorero et al. \cite{Ost10}), 
or synchrotron photons by a dominant jet component 
(Migliori et al., in preparation). 
Lobes' IC emission should be most likely dominant in GPS galaxies,
where projection effects should be marginal,
while IC emission from jet components is favoured 
in more powerful GPS and CSS quasars. 
In the former case, the intensity of the IC emission for a given
jet kinetic power, strictly depends on both the lobe size 
(i.e. its compactness), and the thermal photons arising from the
dense and inhomogeneous nuclear ambient medium embedding the radio
source. In the latter case, the ingredients at the basis of the 
high-energy emission are both the velocity of the emitting region, namely a
jet knot, and its direction (approaching or moving away) 
relative to the source of the external
thermal/non-thermal photon seeds. 
We note that while lobes' IC emission is isotropic, 
the emission from the jet can be strongly beamed. 
More complexity is added when we consider a jet with a velocity
structure, either axial (see e.g. Ghisellini et
al. \cite{ghisellini05}) 
or radial (Celotti et al. \cite{Cel01}, Georganopoulos \& Kazanas
\cite{georga03}), and synchrotron photons 
of the fast moving component are up-scattered 
by the electrons in the slower one.

\begin{figure}
\begin{center}
\includegraphics{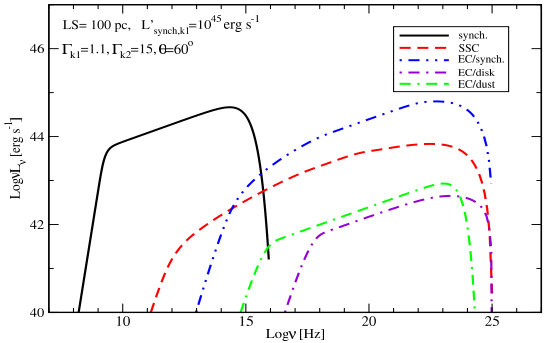}
\vspace{6cm}
\caption{Modeled SED for a knot located at 100 pc from the core (see
  text). The different curves show the modeled components: synchrotron
  emission (synch., {\it black solid line}), SSC emission (SSC, {\it
    red dashed line}), comptonized disc and torus photons (EC/disk and
  EC/dust {\it violet dot-dashed line} and {\it green dot-double dashed line}
  respectively), and IC emission of external synchrotron photons from a blazar-like component (EC/synch., {\it blue double-dot-dashed line}).  }
\label{f2}
\end{center}
\end{figure}

\noindent As an example, we show in Figure \ref{f2} the results for 
synchrotron and IC modeled SED for a knot located at 100 pc 
from the nuclear region, emitting an intrinsic integrated luminosity 
$L_{syn.,k1}'=10^{45}$ erg s$^{-1}$. 
The expected IC emission produced by the knot electrons with 
the local synchrotron photons (synchrotron self Compton, SSC),
and the external Compton with UV/disc (EC/disc) and IR/torus
(EC/torus) photons are calculated.   
The bolometric disc luminosity is $L_{disc}'=10^{46}$ erg s$^{-1}$ and
about $10\%$ of the disc luminosity is reprocessed in the torus. 
The knot is moving with a bulk motion $\Gamma_{k1}=1.1$, 
and the electron energy distribution is described by a simple power
law ($N(\gamma)\propto \gamma^{-p}$) with a energy spectral index
$p=2.6$ and extremes $\gamma_{min}=10$ and $\gamma_{max}=10^5$.
A second knot, which releases an intrinsic integrated synchrotron
luminosity of $L_{syn.,k2}'=10^{43}$ erg s$^{-1}$, 
is located at the jet base and it has a highly relativistic motion 
($\Gamma_{k2}=15$). The synchrotron photons coming from this internal 
blazar-like component are up-scattered by the electrons 
in the slow moving outer knot, (EC/synch., see Celotti et
al. \cite{Cel01} for a complete description of the model). 
The jet axis has an inclination of 60 degrees with respect to the 
observer line of sight. The dominant contribution to the high energy
emission is provided by the EC/synch. emission. 
Then, the presence of a velocity gradient along the jet seems to play
a determinant role in the production of the high energy emission.

\subsection{Non-thermal $\gamma$-ray emission}

An interesting aspect of the non-thermal scenarios,
both for lobes in galaxies and jets in quasars, 
is that compact sources are expected to be also 
important $\gamma$-ray emitters. 
Observations in the $\gamma$-ray band could be important for several aspects:
\begin{itemize}

\item they should allow us to discriminate genuinely young/compact
  sources from projected sources/blazars. In particular, simultaneous, 
multiwavelength (radio to $\gamma$-ray) observations carried out
during various epochs, are
a powerful tool to catch the elusive variability typical of blazars;\\

\item they should be a decisive confirmation of the non-thermal
  hypothesis, since high-energy thermal emission is expected to
  rapidly drop in the MeV-GeV energy bands. Anyway, on this purpose it
  is worth noting that at the current \fermilat\ sensitivity only 
the most powerful and near objects may be detected. Certainly
  this introduces a fundamental bias, especially for the quasar class
  in average located at relatively high redshifts.\\

\item $\gamma$-ray emission from the jet could allow us to shed a
  light on the jet dynamics during the initial stages, 
whether there is a single velocity or a more complex structure. 
In this case, the comparison with the giant counterparts detected by 
\fermilat\ could be also important to better understand the general
evolutionary path.  \\

\end{itemize}

\section{Conclusions}

Intrinsically compact radio galaxies represent a high fraction (15\% -
30\%) of the sources selected in flux-limited radio catalogues. Their
compactness together with their two-sided morphology suggests that
their radio emission is still in a young phase and it would likely
evolve into the large edge-brightened radio galaxies. This class of
objects is characterized by their peaked synchrotron spectrum, likely
produced by SSA. Their physical properties indicate that they are in
equipartition conditions with typical values of the magnetic fields
from a few mG to a few hundred mG in the most compact and brightest
components. Such high magnetic fields cause severe energy losses of
the relativistic electrons producing thus steep optically-thin
spectra. The ambient medium enshrouding these radio sources is quite
dense and inhomogeneous, where jet-cloud interaction may take place
and influence the source growth, for example impeding the jet
expansion. \\
Although these sources are mainly studied in the radio band, their X-ray
emission has been detected with the advent of \Cha and \XMM
observatories. However, the insufficient spatial resolution could not
enable us to constrain the site of the X-ray emission, since both
thermal and non-thermal models are degenerate in the X-ray
band. Non-thermal inverse Compton produced by relativistic electrons
from lobes or jets components with either thermal photon seeds from
the ambient medium (i.e. disc/torus), or non-thermal synchrotron
photons from a relativistic jet knot, may be at the origin of the
high-energy emission. \fermilat\ observations in the $\gamma$-ray
regime will help us in defining the actual region responsible for the
high-energy emission, as well as the structure of the newly born radio
jets. 

\begin{acknowledgements}
This workshop has been supported by the European 
Community Framework Programme 7, Advanced Radio Astronomy in Europe, 
grant agreement no.: 227290.
\end{acknowledgements}

\end{document}